\documentstyle[twocolumn,prb,aps,epsfig,subeqnar,pra]{revtex}

\begin{document}

\title{Tunable tunneling: An application of stationary states of Bose-Einstein condensates in traps of finite depth}

\author{L. D. Carr$^{1}$\cite{byline}, K. W. Mahmud$^{1}$, W. P. Reinhardt$^{1,2}$\\}
\address{$^{1}$Department of Physics, University of Washington, Seattle, WA 98195-1560, USA\\}
\address{$^{2}$Department of Chemistry, University of Washington, Seattle, WA 98195-1700, USA\\}

\date{\today}

\maketitle
\begin{abstract}
The fundamental question of how Bose-Einstein condensates tunnel into a barrier is addressed.  The cubic nonlinear Schr\"odinger equation with a finite square well potential, which models a Bose-Einstein condensate in a quasi-one-dimensional trap of finite depth, is solved for the complete set of localized and partially localized stationary states, which the former evolve into when the nonlinearity is increased.   An immediate application of these different solution types is tunable tunneling. Magnetically tunable Feshbach resonances can change the scattering length of certain Bose-condensed atoms, such as  $^{85}$Rb, by several orders of magnitude, including the sign, and thereby also change the mean field nonlinearity term of the equation and the tunneling of the wavefunction.  We find both linear-type localized solutions and uniquely nonlinear partially localized solutions where the tails of the wavefunction become nonzero at infinity when the nonlinearity increases. The tunneling of the wavefunction into the non-classical regime and thus its localization therefore becomes an external experimentally controllable parameter.
\end{abstract}


\section{Introduction}
\label{sec:intro}

One of the most attractive aspects of trapped dilute gas Bose-Einstein condensates (BEC's) is the number of experimental parameters which can be varied over a wide range.  For example, the effective dimensionality of a condensate can be changed by varying the confining frequencies of a harmonic trapping potential~\cite{petrov1,petrov2}.  The size of nonlinear collective excitations such as solitons~\cite{denschlag1,burger1} and vortices~\cite{matthews1,madison1}, which is characterized by the healing length of the condensate~\cite{dalfovo1}, can be varied over many orders of magnitude by changing the density~\cite{carr26} or using a magnetically tunable Feshbach resonance to change the effective atomic interaction strength~\cite{vogels1}.  The mean field of the BEC is characterized by a macroscopic wavefunction.  It is here demonstrated that, in a trap of finite depth, tunneling and localization of this mean field wavefunction is an externally controllable experimental parameter.


A highly successful mean field model for BEC's is the Gross-Pitaevskii equation~\cite{gross1,pitaevskii1}.  In its \emph{quasi-one-dimensional} regime, which holds when the transverse dimensions of the condensate are on the order of the condensate's healing length and its longitudinal dimension is much longer than its transverse ones~\cite{carr15,carr22} and is the regime considered in this paper, the Gross-Pitaevskii equation is identical to the well-known cubic nonlinear Schr\"odinger equation (NLS).  Conversely, and not considered here, when the condensate's transverse dimensions are much less than the healing length then the Gross-Pitaevskii equation no longer applies and other physical models are required~\cite{petrov2,girardeau1}.

The NLS models many other natural phenomena as well, including spin waves in magnetic materials~\cite{kalinikos1}, Bose-condensed photons~\cite{ciao1}, disordered media~\cite{mamaev1}, helical excitations of a vortex line~\cite{hasimoto1}, and light pulses in optical fibers~\cite{hasegawa1}.  However, as BEC's are trapped, an extra potential term is required.  This has opened up new theoretical and mathematical investigations~\cite{carr15,carr16,carr23,carr24,carr25}.  The stationary cubic NLS with a potential may be written~\cite{carr15}
\begin{equation}
[-\partial_{x}^{2}+\eta\mid\!f(x)\!\mid^{2}+V^{trap}(x)\,]\,f(x) = \mu\,f(x)\, ,
\label{eqn:nls}
\end{equation}
where $f(x)$ is the longitudinal portion of the mean-field condensate wavefunction, $\mu$ is the eigenvalue and the chemical potential, and $\eta$ is the strength of the nonlinearity, an extra parameter from the usual NLS which cannot be scaled out.  All quantities in Eq.~(\ref{eqn:nls}) are dimensionless.  Physical units relevant to present experiments are discussed in Sec.~\ref{sec:application}.  $\eta\propto a N$, where $a$ is the s-wave scattering length which characterizes binary atomic interactions in a dilute gas and $N$ is the number of atoms in the trap.  Note that the transverse degrees of freedom have been frozen out.

In order to consider a BEC in a finite trap the square well potential
\begin{equation}
V^{trap}(x) =
\left\{0\,\,\,\quad |x|\leq l
\atop{V_o\quad |x|>l}\right .
\label{eqn:potential}
\end{equation}
is studied.  This potential gives analytic solutions, unlike harmonic traps, which in one dimension do not give rise to analytic solutions.  It has the advantage of having a direct analog in the linear Sch\"odinger equation, for which the stationary states have been worked out completely.  In addition it models experimental configurations which could be constructed with present BEC technology.  $^{87}$Rb BEC's trapped in hollow blue-detuned laser beams~\cite{bongs1} can be made quasi-one-dimensional~\cite{carr26}.  Laser light sheets already in use as endcaps for such cylindrical traps have a variable intensity $I$ which could make the trap finite.  $^{85}$Rb, for which the interaction length is particularly responsive to magnetic tuning via a Feshbach resonance~\cite{cornish1}, could also be trapped in this manner~\cite{cornellcommunication1}.

Solutions of Eq.~(\ref{eqn:nls}) with the finite square well potential Eq.~(\ref{eqn:potential}) are presented.  When $\eta$ is negative the solutions are localized.  When $\eta$ is positive and the chemical potential $\mu$ is less than or equal to the well depth $V_o$, the solutions are again localized.  However, when $\eta$ is positive and $\mu > V_o$, the solutions are \emph{partially localized}, i.e. the tails of the wavefunction approach a constant non-zero value at infinity.  These latter solutions grow analytically and continuously out of the localized, bound solutions, and have no analog in the linear Schr\"odinger equation.

The transition between a localized and a partially localized stationary state is an adiabatic one which does not change the number of nodes, unlike in the linear case, where the wavefunction aquires nodes when pushed out the top of the well.  The nonlinear analog of scattering states, which continue to oscillate as $x\rightarrow\pm\infty$ and do not grow analytically and continuously out of the bound state solutions, are not considered here.  In the linear Schr\"odinger equation the square well parameters $V_o$ and $l$ completely determine the stationary states, whereas in the NLS $\eta$ is an extra external control parameter which determines the extent of the tunneling and therefore the localization of the wavefunction.  Exploration of this extra freedom is the key advance in this paper.

The mathematical question of how solutions to the NLS decay under a potential step is, to the authors' knowledge, here answered for the first time.  The closest previous treatments have considered a repulsive delta function, which models a small scatterer or impurity~\cite{hakim1}, or approximate solutions to a harmonic potential~\cite{kunze1,turitsyn1,kivshar6}.  Limits of the finite square well include the delta function and the infinite square well, the latter of which has been solved elsewhere~\cite{carr15,carr16}.  Both attractive and repulsive BEC's are treated, i.e. $\eta$ negative and positive.  The problem of double well tunneling, which is well known, in the case of a high barrier, to lead to breakdown of the NLS~\cite{smerzi1,milburn1}, and therefore represents an entirely different physical situation, is not considered.

The article is outlined as follows.  In Sec.~\ref{sec:square} the full set of localized and partially localized stationary solutions of the stationary NLS with a finite square well potential are presented.  In Sec.~\ref{sec:limit} the delta function and infinite square well limits are explicated.  In Sec.~\ref{sec:application} the application of these solutions to tunable tunneling in BEC's is discussed, with specific parameters relevant to current BEC experiments.  Finally in Sec.~\ref{sec:summary} the results are briefly summarized.

\section{Finite square well}
\label{sec:square}

Just as for the linear Schr\"odinger equation with a finite square well potential, solutions to Eq.~(\ref{eqn:nls}) with the potential Eq.~(\ref{eqn:potential}) are constructed by matching decaying solutions outside the well to periodic solutions inside the well, subject to continuity of $f(x)$ and its first derivative at the edges of the well.  Periodic solutions to Eq.~(\ref{eqn:nls}) 
with $V(x)=0$ are constructed from the special functions known as Jacobian elliptic functions~\cite{abramowitz1,bowman1}.  Limits of these functions also constitute decaying solutions.  Elliptic functions are more commonly known as solutions to the anharmonic classical oscillator, i.e. $\ddot{\theta} + \theta - \theta^{3}/3! = 0$.  The standard notation sn$(x \mid m)$ will be used for elliptic functions, where $m$ is the elliptic parameter.  This parameter interpolates the Jacobian elliptic functions between trigonometric and hyperbolic functions.  Their quarter period is given by the complete elliptic integral $K(m)$.  A short summary of the particular functions relevant to this work is given in Table~\ref{table:jacobian}.

Trivial phase solutions of Eq.~(\ref{eqn:nls}) with $V(x)=0$ are the set of twelve elliptic functions~\cite{carr15}.  Of these, six are bounded and six are not.  Of the bounded functions, three have different physical form, i.e. differ by more than a translational shift or a rescaling of the amplitude.  These are cn$(x\mid m)$, dn$(x\mid m)$, and sn$(x\mid m)$, and are all candidates for solutions inside the well.  The $m=1$ hyperbolic limits of the remaining six unbounded functions constitute the tails outside the well.  These are sech$(x)$, csch$(x)$, and coth$(x)$.  $m=1$ corresponds to the infinite wavelength limit of the elliptic functions.  However, only certain combinations of these functions are possible.  For $\eta>0$, sn$(x\mid m)$ is the only periodic solution in the well and coth$(x)$ and csch$(x)$ are both possibilites outside the well.  For $\eta<0$, cn$(x\mid m)$ and dn$(x\mid m)$ are both possible inside the well while sech$(x)$ is the only possible decaying solution outside the well.  Therefore there are four possibilities, two for each sign of $\eta$.  Of these four, two shall be shown to be localized states, one a partially localized state, and one not a solution.  Thus there are three solution types.

All solutions will be written in the form
\begin{equation}
f(x) = \left\{ \begin{array}{lll}
                 f_1(x)\quad & x < -l\\
                 f_2(x)\quad & |x| \leq l\\
                 f_3(x)\quad & x > l
               \end{array}
    \right.
\label{eqn:form}
\end{equation}
in correspondence with the form of the potential, Eq.~(\ref{eqn:potential}).  These will be subject to continuity of $f(x)$ and $f^{\prime}(x)$ at $x=\pm l$ and, for the localized states, the normalization $\int_{-\infty}^{\infty}dx\,|f(x)|^2 = 1$.  For the partially localized states, which are not normalizable, the normalization condition will be replaced with the use of the constant amplitude at infinity as a free parameter.  Note that for the localized states the normalization has been chosen to be 1 rather than the number of atoms $N$ because $N$ has been absorbed into the coefficient $\eta$.  Unlike in linear quantum mechanics, the nonlinear term in the NLS makes the choice of normalization non-trivial.

\subsection{Localized states}
\label{subsec:local}

Localized states are defined as bound states for which the tails approach zero at a large distance from the well.  The functions sech$(x)$ and csch$(x)$ satisfy this criterion and solve the cases of attractive and repulsive nonlinearity, respectively, i.e. $\eta < 0$ and $\eta > 0$.  The fundamental difference between these two functions and $\exp(\pm x)$, which solves the analogous linear Schr\"odinger problem, is not their functional form.  Rather it is an offset $b$ which for attractive nonlinearity retracts the decaying tail into the well and for repulsive nonlinearity pushes the tail into the barrier.  That is, for $x > l$ the functional form of the wavefunction is sech$(x+b)$ with $b < 0$ for the former case and csch$(x+b)$ with $b > 0$ in the latter one.  This can be understood as a physical effect of the nonlinear term in Eq.~(\ref{eqn:nls}).  Note in the limit as $x\rightarrow\pm\infty$, sech$(x+b)$ and $|\text{csch}(x+b)|$ both approach exp$[\mp (x+b)]$.

\subsubsection{Attractive nonlinearity}
\label{subsubsec:att}

Symmetric solutions take the form
\begin{subeqnarray}
\label{eqn:localA1}
  f_1(x)&=& A\,\text{sech}[k(x-b)]\, ,\\
  f_2(x)&=& A_2\,\text{cn}(k_2 x\mid m)\, ,\\
  f_3(x)&=& A\,\text{sech}[k(x+b)]\, ,
\end{subeqnarray}
and antisymmetric solutions take the form
\begin{subeqnarray}
\label{eqn:localA2}
  f_1(x)&=& -A\,\text{sech}[k(x-b)]\, ,\\
  f_2(x)&=& A_2\,\text{cn}[k_2 x + K(m) \mid m]\, ,\\
  f_3(x)&=& A\,\text{sech}[k(x+b)]\, ,
\end{subeqnarray}
where $A$, $k$, $k_2$, $b$, and $m$ are free parameters.  $f_1(x)$ and $f_3(x)$ have been chosen in a manner to preserve symmetry under the reflection operation $x\rightarrow -x$.  In App.~\ref{app:dn} it is proven that the analogous solutions with dn in place of cn in Eq.~(\ref{eqn:localA1}b) or Eq.~(\ref{eqn:localA2}b) cannot simultaneously satisfy normalization and boundary conditions.  The dn elliptic function is a solution type dependent upon special symmetries, as has been exposited elsewhere~\cite{carr16}.  The finite square well breaks these symmetries.

Eqs.~(\ref{eqn:localA1}) and~(\ref{eqn:localA2}) are subject to the same solution method.  Consider first Eqs.~(\ref{eqn:localA1}).  Substituting them into Eq.~(\ref{eqn:nls}) with the potential Eq.~(\ref{eqn:potential}), the conditions
\begin{subeqnarray}
\label{eqn:localA3}
  A^2&=& -2k^2/\eta\\
  \mu&=& V_o - k^2\\
  A_2^2&=& -2mk_2^2/\eta\\
  \mu&=& - (2m-1)k_2^2
\end{subeqnarray}
are obtained.  The derivatives of elliptic functions, used to obtain Eqs.~(\ref{eqn:localA3}), may be found in the literature~\cite{abramowitz1}.  The boundary condition $f_1(-l)=f_2(-l)$ is equivalent to $f_3(l)=f_2(l)$ for Eqs.~(\ref{eqn:localA1}), and requires
\begin{equation}
\label{eqn:localA4}
A\,\text{sech}[k(l+b)]=A_2\,\text{cn}(k_2 l\mid m)\,.
\end{equation}
Continuity of the first derivative requires
\begin{eqnarray}
\label{eqn:localA5}
A k\,\text{tanh}[k(l+b)]\,\text{sech}[k(l+b)]\quad \quad \nonumber\\
=A_2 k_2\,\text{dn}(k_2 l\mid m)\,\text{sn}(k_2 l\mid m)\,.
\end{eqnarray}
Finally normalization requires
\begin{eqnarray}
\label{eqn:localA6}
2A_2^2\int_0^l dx\, \text{cn}^2(k_2 x\mid m)&&\nonumber\\
 + 2A^2\int_l^{\infty} dx\, \text{sech}^2&[k(x+b)] =& 1 \, .
\end{eqnarray}
Using the Jacobian elliptic identity $m\,\text{cn}^2(x\mid m) +(1-m)= \text{dn}^2(x\mid m)$ and the integral relation $\int_0^{K(m)} dx\,\text{dn}^2(x\mid m) = E(m)$, together with the hyperbolic relations $\int_y^{\infty}dy\,\text{sech}^2(y) = 1 - \text{tanh}(y) = \exp(-y)\,\text{sech}(y)$, Eq.~(\ref{eqn:localA6}) can be written in the form
\begin{eqnarray}
\label{eqn:localA7}
\frac{2 A_2^2}{k_2 m}\,[E(k_2 l\mid m)-(1-m)k_2 l]&\nonumber \\
+\frac{2A^2}{k}\,\exp[-k(l+b)]\,\text{sech}[k(&l+b)] = 1 \, .
\end{eqnarray}
where $E(k_2 l\mid m)$ is standard notation for an incomplete elliptic integral~\cite{abramowitz1}.

Equation of Eqs.~(\ref{eqn:localA3}b) and~(\ref{eqn:localA3}d) and substitution of Eqs.~(\ref{eqn:localA3}a) and~(\ref{eqn:localA3}c) into Eqs.~(\ref{eqn:localA4}),~(\ref{eqn:localA5}), and ~(\ref{eqn:localA7}) produce a system of four simultaneous equations:
\begin{subeqnarray}
\label{eqn:localA8}
&(1-2m)k_2^2 l^2 + k^2 l^2 = V_o l^2\, ,\\
&\nonumber\\
&k\,l \,\text{sech}[k(l+b)]=\sqrt{m}\,k_2 l\,\text{cn}(k_2 l\mid m)\, ,\\
&\nonumber\\
&k^2 l^2\,\text{tanh}[k(l+b)]\,\text{sech}[k(l+b)]\nonumber\\
&=\sqrt{m}\,k_2^2 l^2\,\text{dn}(k_2 l\mid m)\,\text{sn}(k_2 l\mid m)\, ,\\
&\nonumber\\
&k_2\,l\,[E(k_2 l\mid m)-(1-m)k_2 l]\nonumber \\
&+k\, l\,\exp[-k(l+b)]\,\text{sech}[k(l+b)] = -l\eta/4 \, .
\end{subeqnarray}
All equations have been multiplied through by an extra factor of $l$.  A notational simplification can be made by defining $\alpha\equiv k\,l$, $\beta\equiv k_2 l$, $\omega\equiv k(l+b)$, and $\gamma = \sqrt{V_o}l$.  Then
\begin{subeqnarray}
\label{eqn:localA9}
&(1-2m)\beta^2 + \alpha^2 = \gamma^2\, ,\\
&\nonumber\\
&\alpha\,\text{sech}(\omega)=\sqrt{m}\,\beta\,\text{cn}(\beta\mid m) \, ,\\
&\nonumber\\
&\alpha^2\,\text{tanh}(\omega)\,\text{sech}(\omega)\nonumber\\
&=\sqrt{m}\,\beta^2\,\text{dn}(\beta\mid m)\,\text{sn}(\beta\mid m) \, ,\\
&\nonumber\\
&\beta\,[E(\beta\mid m)-(1-m)\beta]+\alpha\,e^{-\omega}\,\text{sech}(\omega)\nonumber \\ 
&= -l\eta/4 \, .
\end{subeqnarray}
These four equations are in four unknowns, $\alpha$, $\beta$, $\omega$, and $m$, with an additional three variables $l$, $\eta$, and $\gamma$ which are externally determined experimental parameters.

This system of equations may be compared to the analogous set derived for the linear Schr\"odinger equation, as may be found in any undergraduate quantum mechanics textbook~\cite{bransden1}.  In the linear case, the energy condition 
Eq.~(\ref{eqn:localA8}a) describes a circle.  Here it describes a hyperbola for $1-2m<0$ and an ellipse for $1-2m>0$.  Equations~(\ref{eqn:localA8}b) and~(\ref{eqn:localA8}c), derived from boundary conditions on $f(x)$, are similar in form to the linear ones, which are usually combined by dividing one by the other.  Finally Eq.~(\ref{eqn:localA8}d) has no linear analog, as in the linear case normalization is trivial, and therefore here forms an extra condition.  The linear system is reduced to two transcendental equations and solved by graphical methods.  Here a similiar approach is used.

\begin{figure}
\begin{center}
\epsfig{figure=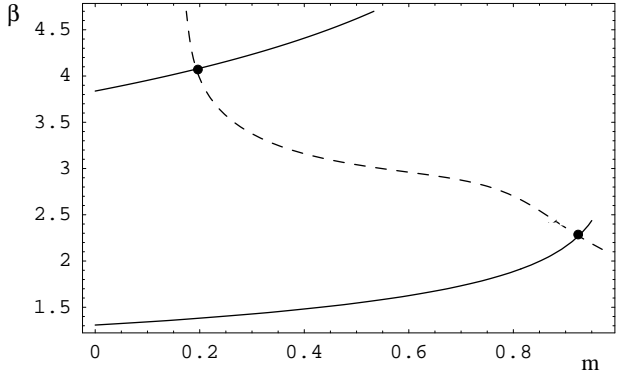,width=8.2cm}
\end{center}
\caption{
Graphical solutions method:  shown are the first two intersections of Eqs.~(\ref{eqn:localA10}a) (solid line) and~(\ref{eqn:localA10}b) (dashed line.)  These give the ground state and first symmetric excited state, at the lower right and upper left of the plot, respectively.
}
\label{fig:2}
\end{figure}

Two unknowns may be eliminated as follows.  Equations~(\ref{eqn:localA9}a) and~(\ref{eqn:localA9}b) are solved for $\alpha$ and $\omega$.  These are then substituted into Eqs.~(\ref{eqn:localA9}c) and~(\ref{eqn:localA9}d) to obtain
\begin{subeqnarray}
\label{eqn:localA10}
&\sqrt{m}\,\text{tanh}\{\text{sech}^{-1}\left[\lambda\,\text{cn}(\beta\mid m)\right ]\}\nonumber \\
&=\lambda\,\text{dn}(\beta\mid m)\,\text{sc}(\beta\mid m)\, , \\
&\nonumber\\
&\sqrt{m}\,\beta\,\text{cn}(\beta\mid m)\,\text{exp}(-\{\text{sech}^{-1}[\lambda\,\text{cn}(\beta\mid m)]\})\nonumber\\
&+\beta\,[E(\beta\mid m)-(1-m)\beta ]=-l\eta/4\, ,
\end{subeqnarray}
where $\lambda\equiv\sqrt{m}\,\beta/\sqrt{\gamma^2-(1-2m)\beta^2}$.  The resulting two equations in the two unknowns $\beta$ and $m$ may be solved implicitly by a combination of graphical and numerical techniques.  The method is demonstrated in Fig.~\ref{fig:2}.  The approximate location of the solution in parameter space is first determined by graphical examination of the intersection of Eqs.~(\ref{eqn:localA10}).   Then the well-known secant method is used to find $(m,\beta)$ to double precision accuracy.  The largest value of $m$ for which Eqs.~(\ref{eqn:localA10}) intersect gives the ground state and the excitation levels are thereafter monotonically decreasing in $m$.  In Fig.~\ref{fig:2}, a well of depth $V_o=25$ and width $2l=2$ and a nonlinearity of $\eta=-8$ were used. The intersection in the lower right corner of the plot, at $m=0.9258$ and $\beta=2.2709$, gives the ground state.  The other illustrated intersection, at $m=0.1934$ and $\beta=4.0749$, gives the first symmetric excited state.

\begin{figure}
\begin{center}
\epsfig{figure=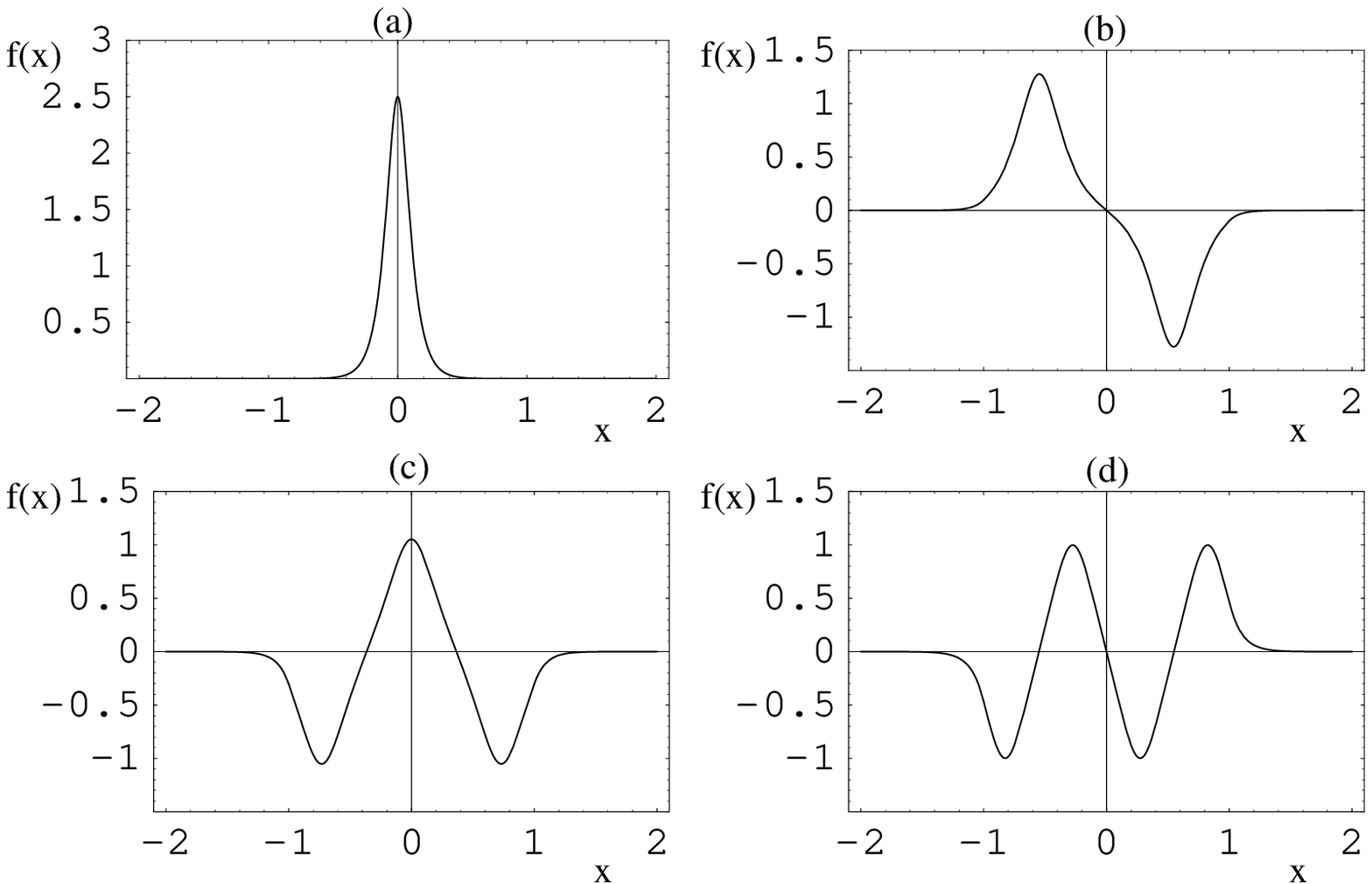,width=8.2cm}
\end{center}
\caption{
Shown are the ground state and first three excited states for localized solutions with attractive nonlinearity $\eta$.  The box walls are at $x=\pm 1$.
}
\label{fig:1}
\end{figure}

The antisymmetric case Eqs.~(\ref{eqn:localA2}) may be treated with 
a similiar solution method.  The system of equations
\begin{subeqnarray}
\label{eqn:localA11}
&(1-2m)\beta^2 + \alpha^2 = \gamma^2\, ,\\
&\nonumber \\
&\alpha\,\text{sech}(\omega)=\beta\,\text{cn}[\beta+K(m)\mid m] \, ,\\
&\nonumber\\
&\alpha^2\,\text{tanh}(\omega)\,\text{sech}(\omega)\nonumber\\
&=\beta^2\,\text{dn}[\beta+K(m)\mid m)]\,\text{sn}[\beta+K(m)\mid m] \, ,\\
&\nonumber\\
&\beta\,\{E[\beta+K(m)\mid m]-(1-m)\beta\}\nonumber \\ 
&+\alpha\,e^{-\omega}\,\text{sech}(\omega)= -l\eta/4
\end{subeqnarray}
results.  The ground state and first three excited states, two of which are symmetric and two of which are antisymmetric, are plotted in Fig.~\ref{fig:1} for $\eta= -50$, $V_o=100$, and $2l=2$.  For the ground state, the Jacobi elliptic parameter $m = 0.999999999972$ and the chemical potential is $\mu=-156.25$.  For the first three excited states, $m$ and $\mu$ are 0.9857, 0.7694, 0.5318 and -40.20, -19.37, and -2.982, respectively.  Higher excited states are closer to linear Schr\"odinger equation, sinusoidal-type solutions, while lower states are visibly nonlinear. The ground state become more bound, i.e. the eigenvalue becomes more negative, as the absolute value of the nonlinearity increases.  Atoms tend to cluster at the center of the well because of the attractive nonlinearity, as can be seen in Fig.~\ref{fig:1}(a).

\subsubsection{Repulsive nonlinearity}
\label{subsubsec:rep}

Symmetric solutions take the form
\begin{subeqnarray}
\label{eqn:localB1}
  f_1(x)&=& -A\,\text{csch}[k(x-b)]\, ,\\
  f_2(x)&=& A_2\,\text{sn}[k_2 x+K(m)\mid m]\, ,\\
  f_3(x)&=& A\,\text{csch}[k(x+b)]\, ,
\end{subeqnarray}
and antisymmetric solutions take the form
\begin{subeqnarray}
\label{eqn:localB2}
  f_1(x)&=& A\,\text{csch}[k(x-b)]\, ,\\
  f_2(x)&=& A_2\,\text{sn}(k_2 x\mid m)\, ,\\
  f_3(x)&=& A\,\text{csch}[k(x+b)]\, ,
\end{subeqnarray}
where amplitudes and offsets have again been chosen to preserve symmetry under reflection.  Substitution of these equations into Eq.~(\ref{eqn:nls}) with the potential Eq.~(\ref{eqn:potential}) determines the amplitudes and chemical potential
\begin{subeqnarray}
\label{eqn:localB3}
  A^2&=& 2k^2/\eta\, ,\\
  A_2^2&=& 2mk_2^2/\eta\, ,\\
  \mu&=&(1+m)k_2^2=V_0-k^2\, ,
\end{subeqnarray}
and, together with boundary conditions, the simultaneous transcendental equations which determine the solutions up to the three external experimental parameters $\eta$, $l$, and $\gamma$ are
\begin{subeqnarray}
\label{eqn:localB4}
&(1+m)\beta^2 + \alpha^2 = \gamma^2\, ,\\
&\nonumber \\
&\alpha\,\text{csch}(\omega)=\sqrt{m}\,\beta\,\text{sn}[\beta+K(m)\mid m] \, ,\\
&\nonumber\\
&\alpha^2\,\text{coth}(\omega)\,\text{csch}(\omega)=-\sqrt{m}\,\beta^2\,\text{dn}[\beta+K(m)\mid m]\nonumber\\
&\times\,\text{cn}[\beta+K(m)\mid m] \, ,\\
&\nonumber\\
&\beta\,\{\beta-E[\beta+K(m)\mid m]+E(m)\}\nonumber \\ 
&+\alpha\,e^{-\omega}\,\text{csch}(\omega)= l\eta/4 \, ,
\end{subeqnarray}
for the symmetric case.  A similiar set is obtained for the antisymmetric case.

\begin{figure}
\begin{center}
\epsfig{figure=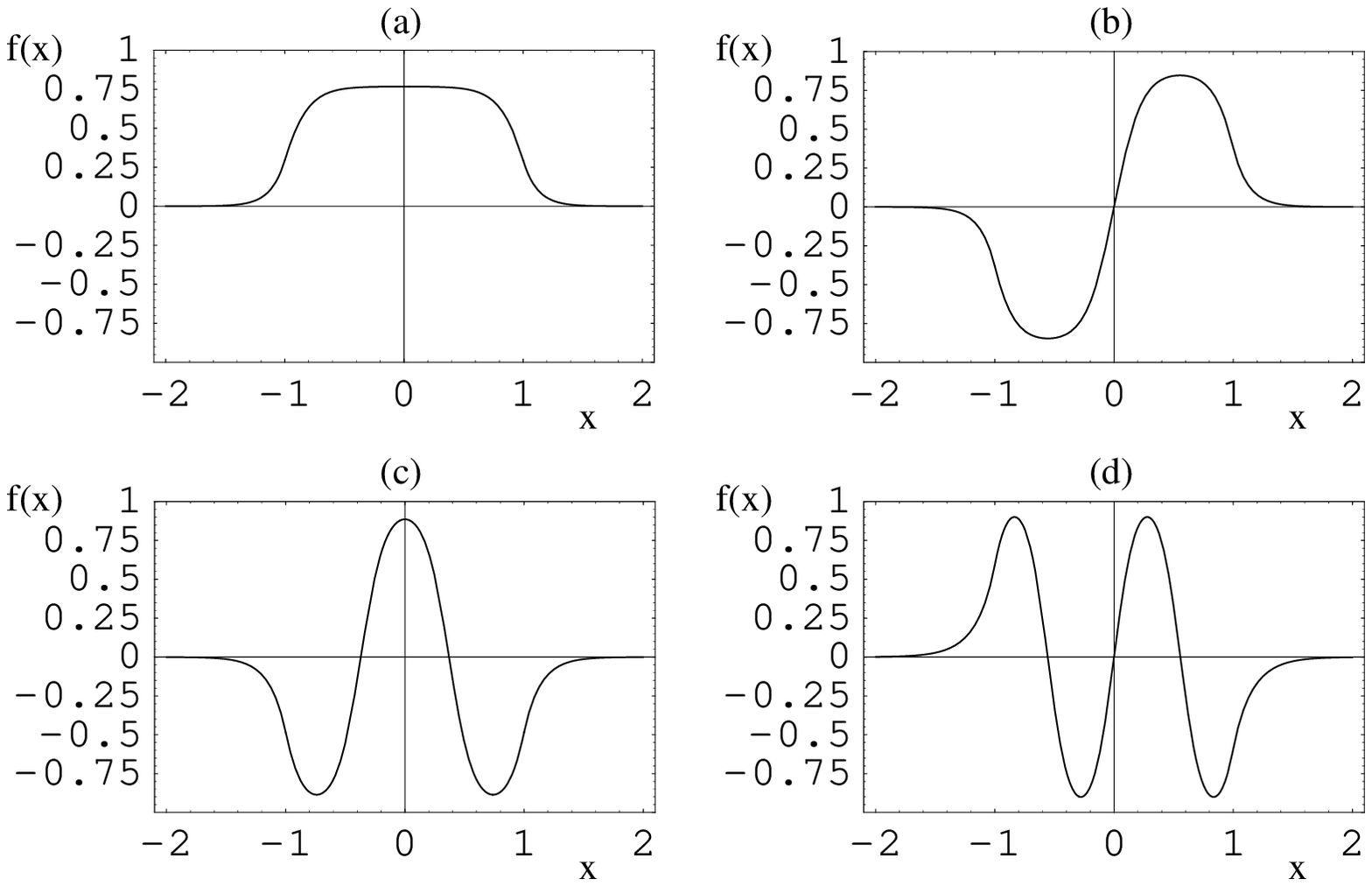,width=8.2cm}
\end{center}
\caption{
Shown are the ground state and first three excited states for localized solutions with positive nonlinearity $\eta$.  The box walls are at $x=\pm 1$.
}
\label{fig:3}
\end{figure}

The ground state and first three excited states are plotted in Fig.~\ref{fig:3} for $\eta = 50$ and the same well dimensions as were used in Fig.~\ref{fig:1}, $V_o = 100$ and $2l = 2$.  In panels (a)-(d) the elliptic parameter $m$ and chemical potential $\mu$ are 0.9967, 0.8808, 0.6588, 0.4696 and 29.47, 38.23, 49.40, 63.45, respectively. As the nonlinearity is increased the ground state chemical potential increases until it reaches the height of the well for $\eta\sim 210$. For $\eta\gtrsim 210$ there are no normalizable localized bound states in the potential well.  The solution then becomes partially localized, as described in the next section.          

\subsection{Partially localized states}
\label{subsec:partial}

Partially localized states are defined as states for which the tails approach a constant non-zero value at a large distance from the well but the density $|f(x)|^2$ is still higher in the region of the well.  They have no analog in the linear Schr\"odinger equation.  As stated in Sec.~\ref{sec:intro}, the analog of scattering states, which continue to oscillate as $x\rightarrow\pm\infty$, are not considered.  Rather, solutions which follow the adiabatic evolution of the localized, bound states described above, i.e. those which continue to have the same number of nodes as $\eta$ is increased, are presented.

The ground state of the NLS with $V(x)=0$ and $\eta>0$ is a constant, $f(x)=\sqrt{\mu}$.  Partially localized states can
be understood as a deformation of this ground state: the wavefunction spills out over the top of the well and approaches a constant non-zero value as $x\rightarrow\pm\infty$.  If the dispersion term in the NLS is neglected, in the so-called Thomas-Fermi limit~\cite{dalfovo1}, the ground state solution is $f(x)=\sqrt{\mu-V(x)}$ inside the well.  If $\mu \leq V_o$ the wavefunction is zero outside the well; if $\mu > V_o$ then it is a non-zero constant.  The ground state analytic solution presented here is, in some regimes, similiar to that of the Thomas-Fermi limit.  But it gives the correct, continuous connection between the wavefunction inside and outside the well and, unlike the Thomas-Fermi limit, describes the excited states as well.

\begin{figure}
\begin{center}
\epsfig{figure=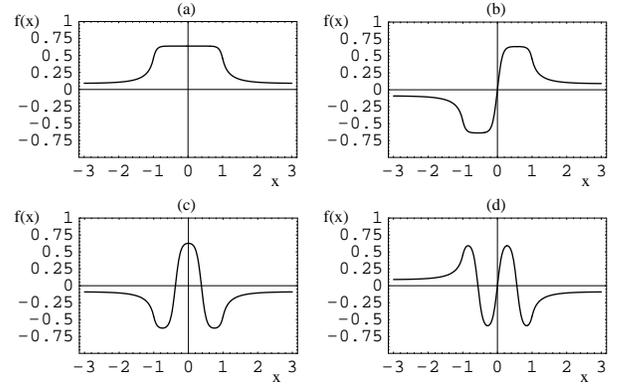,width=8.2cm}
\end{center}
\caption{
Shown are the ground state and first three excited states for partially localized solutions, for which the tails of the wavefunction approach a non-zero value as $x\rightarrow\pm\infty$.  The box walls are at $x=\pm 1$.  Here the tails of the wavefunction have been chosen to approach $\frac{1}{11}$.
}
\label{fig:4}
\end{figure}

Symmetric solutions take the form
\begin{subeqnarray}
\label{eqn:localC1}
  f_1(x)&=& -A\,\text{coth}[k(x-b)]\, ,\\
  f_2(x)&=& A_2\,\text{sn}[k_2 x + K(m) \mid m]\, ,\\
  f_3(x)&=& A\,\text{coth}[k(x+b)]\, ,
\end{subeqnarray}
and antisymmetric solutions take the form
\begin{subeqnarray}
\label{eqn:localC2}
  f_1(x)&=& A\,\text{coth}[k(x-b)]\, \\
  f_2(x)&=& A_2\,\text{sn}(k_2 x\mid m)\, \\
  f_3(x)&=& A\,\text{coth}[k(x+b)]\, .
\end{subeqnarray}
The amplitudes and chemical potential are
\begin{subeqnarray}
\label{eqn:localC3}
  A^2&=& 2k^2/\eta\, ,\\
  A_2^2&=& 2mk_2^2/\eta\, ,\\
  \mu&=&(1+m)k_2^2=V_0+2k^2\, ,
\end{subeqnarray}
while the set of simultaneous transcendental equations are
\begin{subeqnarray}
\label{eqn:localC4}
&(1+m)\beta^2 -2 \alpha^2 = \gamma^2\, ,\\
&\nonumber \\
&\alpha\,\text{coth}(\omega)=\sqrt{m}\,\beta\,\text{sn}[\beta+K(m)\mid m] \, ,\\
&\nonumber\\
&\alpha^2\,\text{csch}^2(\omega)=-\sqrt{m}\,\beta^2\,\text{dn}[\beta+K(m)\mid m]\nonumber\\
&\times\,\text{cn}[\beta+K(m)\mid m] \, ,
\end{subeqnarray}
for the symmetric case.  A similiar set is obtained for the antisymmetric case.  For partially localized states there is no normalization condition.  Instead $A$ is a free parameter.  The ground state and three excited states are shown in Fig.~\ref{fig:4} for $\eta = 250$, the same well dimensions as in Figs.~\ref{fig:1} and~\ref{fig:3}, and $k = 1$.  The parameter $m$ in panels (a)-(d) is 0.9999983, 0.9948, 0.9272 and 0.7466, respectively. Here $k$ is a free parameter that determines the chemical potential and the amplitude $A$. $A$ is also determined by $\eta$, as is apparent in Eq.~(\ref{eqn:localC3}a).

\subsection{Transition from localized to partially localized states}
\label{subsec:transition}

As mentioned in Sec.~\ref{sec:intro}, localized and partially localized states are differentiated by the chemical potential.  For $\eta > 0$, $\mu \leq V_o$ results in the former solution type and $\mu > V_o$ in the latter one.  It is therefore natural to verify that they converge in the limit in which $\mu \rightarrow V_o$.  It is once more emphasized that the transition is an adiabatic, continuous one which fixes the number of nodes.


Consider first the limit of the partially localized states.  $\mu\rightarrow V_o^+$ implies that $A\rightarrow 0$, which in turn implies $k\rightarrow 0$.  Recall that $\alpha\equiv k\,l$ and $\omega\equiv k(l+b)$.  This implies that the arguments of the hyperbolic functions in Eqs.~(\ref{eqn:localC4}) approach zero.  In the limit as $y\rightarrow 0$, coth$(y)\rightarrow 1/y$ and csch$(y)\rightarrow 1/y$.  Therefore $\alpha\,\text{coth}(\omega)\rightarrow l/(l+b)$, $\alpha^2\text{csch}^2(\omega)\rightarrow [l/(l+b)]^2$, and to zeroth order Eqs.~(\ref{eqn:localC4}) become
\begin{subeqnarray}
\label{eqn:localC5}
&(1+m)\beta^2 = \gamma^2\, ,\\
&\nonumber \\
&l/(l+b)=\sqrt{m}\,\beta\,\text{sn}[\beta+K(m)\mid m] \, ,\\
&\nonumber\\
&[l/(l+b)]^2=-\sqrt{m}\,\beta^2\,\text{dn}[\beta+K(m)\mid m]\nonumber\\
&\times\,\text{cn}[\beta+K(m)\mid m] \, .
\end{subeqnarray}

Consider next the limit of the localized states presented in Sec.~\ref{subsubsec:rep}, $\mu\rightarrow V_o^-$.  In this case Eq.~(\ref{eqn:localB3}c) requires $k\rightarrow 0$.  Then by the same reasoning as was used for the partially localized state limit above, to zeroth order Eqs.~\ref{eqn:localB4} become
\begin{subeqnarray}
\label{eqn:localC6}
&(1+m)\beta^2 = \gamma^2\, ,\\
&\nonumber \\
&l/(l+b)=\sqrt{m}\,\beta\,\text{sn}[\beta+K(m)\mid m] \, ,\\
&\nonumber\\
&[l/(l+b)]^2=-\sqrt{m}\,\beta^2\,\text{dn}[\beta+K(m)\mid m]\nonumber\\
&\times\,\text{cn}[\beta+K(m)\mid m] \, ,\\
&\nonumber \\
&\beta\,\{\beta-E[\beta+K(m)\mid m]+E(m)\}\nonumber \\ 
&+l/(l+b)= l\eta/4 \, .
\end{subeqnarray}
The sets of equations~(\ref{eqn:localC5}) and~(\ref{eqn:localC6}) are identical up to the normalization Eq.~(\ref{eqn:localC6}d), and, as $A\rightarrow 0$ for the partially localized states in this limit, normalization may be imposed.  

The two solution types are also distinct in their nonlinearity $\eta$, i.e. there is an $\eta_0$ for which a transition occurs for a particular set of well parameters and excitation level.  This provides a simple physical interpretation, as is discussed in Sec.~\ref{sec:application}.  Equations~(\ref{eqn:localC6}) can be solved implicitly for $m$ and $\eta$ when reduced to two equations of the form
\begin{subeqnarray}
\label{eqn:localC7}
&\, \sqrt{m}=-\text{ds}[\lambda+K(m)\mid m]\,\text{cs}[\lambda+K(m)\mid m]   \, ,\\
&\nonumber\\
&\eta_0 = \frac{4\lambda}{l}\,\left\{\lambda-E[\lambda+K(m)\mid m]+E(m)\right . \nonumber \\ 
& +\sqrt{m}\,\text{sn}[\lambda+K(m)\mid m]\}  \, ,
\end{subeqnarray}
where $\lambda\equiv\gamma/\sqrt{1+m}$.  In practice, $m$ may be as singularly close to 1 as $m=1-10^{-60}$, especially for deep wells.  Equations~(\ref{eqn:localC7}) may be solved for any excitation level by first finding the roots of Eq.~(\ref{eqn:localC7}a) in descending order and then using the resulting values of $m$ to determine $\eta_0$ from Eq.~(\ref{eqn:localC7}b), with care given to excluding extraneous roots.

To find the transition for the ground state the appropriate simplification is to expand Eqs.~(\ref{eqn:localC7}a) and~(\ref{eqn:localC7}b) to first order in $m_1\equiv 1-m$, solve for $m_1$ in the first equation, and substitute the result into the second.  By this method the ground state transition occurs for
\begin{equation}
\label{eqn:transitioneta}
\eta_0 = \frac{\gamma}{l}\left [ \sqrt{2}\left( \frac{e^{-\sqrt{2}\gamma}}{1+2e^{-\sqrt{2}\gamma}-e^{-2\sqrt{2}\gamma}}\right )+2\gamma\right ]\, .
\end{equation}
In the limit for which $\gamma \gg 1$, i.e. for a deep well, $\eta_0$ further reduces to
\begin{equation}
\label{eqn:transitioneta2}
\eta_0 = \frac{1}{l}\left (2\gamma^2 + 2\sqrt{2}\gamma-\gamma e^{-\sqrt{2}\gamma}\right)\, ,
\end{equation}
which to lowest order is simply $\eta_0=2 V_0$.  For typical experimental parameters, as discussed in Sec.~\ref{sec:application}, the error in this approximation is negligible.

Because determining $\eta_0$ requires numerical methods and/or approximations, $\mu$ is a better analytical quantity to characterize the dividing line between the localized and partially localized states.  Then each of the three solution types exists in a separate regime:  the localized states of Sec.~\ref{subsubsec:att} for $\eta<0$; the localized states of Sec.~\ref{subsubsec:rep} for $\eta > 0$ and $\mu \leq V_o$; and the partially localized states for $\eta> 0$ and $\mu > V_o$.  Note that the case of $\eta = 0$ is that of the linear Schr\"odinger equation.

\section{Limiting cases}
\label{sec:limit}

Two limiting cases are presented.  In Sec.~\ref{subsec:delta} it is demonstrated that the finite square well potential gives the same localized and partially localized solutions as the delta function potential
\begin{equation}
\label{eqn:delta}
V(x)=\zeta\,\delta(x)
\end{equation}
in the limit $l\rightarrow  0$ and $V_o \rightarrow \infty$ with the area of the square well, $2 V_o\, l $, held constant.  Eq.~(\ref{eqn:delta}) models a fixed impurity or scatterer of a size much smaller than the condensate healing length.  In Sec.~\ref{subsec:infinite} it is shown that the limit $V_o \rightarrow \infty$ reproduces the full set of stationary states of the NLS under box boundary conditions~\cite{carr15,carr16}.  Other limits, such as $\eta\rightarrow 0$ or $l \rightarrow \infty$, are not worked out explicitly but reproduce the correct solutions to those cases.

\subsection{Delta function}
\label{subsec:delta}

Consider the delta function limit of the localized, bound solutions presented in Sec.~\ref{subsubsec:att}.  In order for the limit to be tractable,  the square well potential of Eq.~(\ref{eqn:potential}) must be shifted by $V_o$, so that the potential is everywhere zero except between $l$ and $-l$, where it is $-V_o$, and $\mu=-k^2$ outside the well.  Let $l\rightarrow  0$ and $V_o \rightarrow \infty$ with $2 V_o\, l \equiv \kappa$ held constant in Eqs.~(\ref{eqn:localA8}).  Equations~(\ref{eqn:localA8}b) -~(\ref{eqn:localA8}d) then reduce to
\begin{subeqnarray}
\label{eqn:delta8}
&k\,\text{sech}(k\,b)=\sqrt{m}\,k_2\, ,\\
&\nonumber\\
&k^2 \,\text{tanh}(k\,b)\,\text{sech}(k\,b)=\sqrt{m}\,k_2^2 \,(k_2\,l)\, ,\\
&\nonumber\\
&k\, \exp(-k\,b)\,\text{sech}(k\,b) = -\eta/4 \, ,
\end{subeqnarray}
where sn$(k_2\,l\mid m)\sim k_2\, l$ has been used.  Dividing Eq.~(\ref{eqn:delta8}b) by Eq.~(\ref{eqn:delta8}a), solving for $k_2\,l$, and substituting it into Eq.~(\ref{eqn:localA8}a) with $m\rightarrow 1$ results in
\begin{equation}
\label{eqn:delta9}
2k\,\text{tanh}(k\,b)=-\kappa \, .
\end{equation}

If one makes the identification $\kappa = \zeta$ then Eq.~(\ref{eqn:delta9}), together with Eq.~(\ref{eqn:delta8}c),  correctly describes the localized solution to Eq.~(\ref{eqn:nls}) with the potential Eq.~(\ref{eqn:delta}) of the form
\begin{equation}
\label{eqn:delta1}
f(x) = \left\{ \begin{array}{ll}
               A\,\text{sech}[k(x-b)]  \quad & x \leq 0\\
               A\,\text{sech}[k(x+b)]  \quad & x \geq 0
               \end{array}
    \right.
\end{equation}
for $\eta<0$.  Equation~(\ref{eqn:delta9}) is the delta function boundary condition while Eq.~(\ref{eqn:delta8}c) is the normalization, as the reader may verify.  $k$ and $b$ can be solved for explicitly to yield  
\begin{subeqnarray}
\label{eqn:delta7}
&b=\frac{-2\log_e(1+4\zeta/\eta)}{\eta+2\zeta}\, ,\\
&\nonumber \\
&k=-\frac{\zeta}{2}-\frac{\eta}{4} \, .
\end{subeqnarray}
The amplitude and chemical potential are the same, namely $A^2=-2k^2/\eta$ and $\mu=-k^2$.  Note that in the limit in which $\eta\rightarrow 0$ the linear Schr\"odinger equation is recovered, $k=-\zeta/2$, and $\mu=\zeta^2/4$, which reproduces the eigenvalue of the well-known linear solution.

The delta function limit of the solutions presented in Sec.~\ref{subsubsec:rep} and Sec.~\ref{subsec:partial} is taken in a similiar manner.  These are similiar to Eq.~(\ref{eqn:delta1}) but have the functional form $f(x)=\pm A\,\text{csch}[k(x\pm b)]$ and $f(x)=\pm A\,\text{tanh}[k(x\pm b)]$ for localized and partially localized solutions, respectively.  Non-trivial phase solutions, which are essentially captured solitons, have been treated by Hakim~\cite{hakim1}.  Note also that $A\,\text{tanh}(k\,x)$ is a solution since it has a node at the delta function, as well as various Jacobian elliptic functions which also have a node at $x=0$.  These cases are not treated here.

\subsection{Infinite square well}
\label{subsec:infinite}

It shall be demonstrated that the localized solutions of Sec.~\ref{subsec:local} reproduce the infinite square well solutions~\cite{carr15,carr16} in the limit in which $V_o\rightarrow\infty$.  Partially localized states have no such limit, as they cannot satisfy the boundary conditions.

Consider first the localized solutions of Sec.~\ref{subsubsec:att}.  $V_o\rightarrow\infty$ implies that $k\rightarrow\infty$.  Then Eqs.~(\ref{eqn:localA4}) and~(\ref{eqn:localA8}) reduce to 
\begin{subeqnarray}
\label{eqn:infty1}
&A_2^2= -2mk_2^2/\eta\, ,\\
&\nonumber\\
&\mu= - (2m-1)k_2^2\, ,\\
&\nonumber\\
&0=\sqrt{m}\,k_2 l\,\text{cn}(k_2 l\mid m)\, ,\\
&\nonumber\\
&k_2\,l\,[E(k_2 l\mid m)-(1-m)k_2 l]=-l\eta/4 \, .
\end{subeqnarray}
The same must hold true for $l\rightarrow -l$.  Equation~(\ref{eqn:infty1}c) then requires that the argument of the elliptic function cn be an integer number of half periods, i.e. $k_2 l = (2p+1)K(m)$, where $p$ is an integer.  $E[(2p+1)K(m)\mid m]=(2p+1)E(m)$, from which it follows that the amplitude, chemical potential, and normalization condition are
\begin{subeqnarray}
\label{eqn:infty2}
&A_2^2= -2m[(2p+1)K(m)]^2/\eta\, ,\\
&\nonumber\\
&\mu= - (2m-1)[(2p+1)K(m)]^2\, ,\\
&\nonumber\\
&(2p+1)^2 K(m)\,[E(m)-(1-m)K(m)]=-\frac{\eta}{4}\, .
\end{subeqnarray}
In the notation of reference [20], $-\eta=2\lambda^{-2}$ and $\tilde{\mu}=-\mu/\eta$.  $2j=2p+1$, where $j$ is an integer, since the edges of the well shift from $-l,l$ to $0,2l$.  With these identifications Eqs.~(\ref{eqn:infty2}) correctly reproduce the infinite square well solutions for attractive nonlinearity~\cite{carr16}.  For repulsive nonlinearity~\cite{carr15}, the limit of the localized solutions treated in Sec.~\ref{subsubsec:rep} is found in a similar manner.

\section{Application: tunable tunneling}
\label{sec:application}

\begin{figure}
\begin{center}
\epsfig{figure=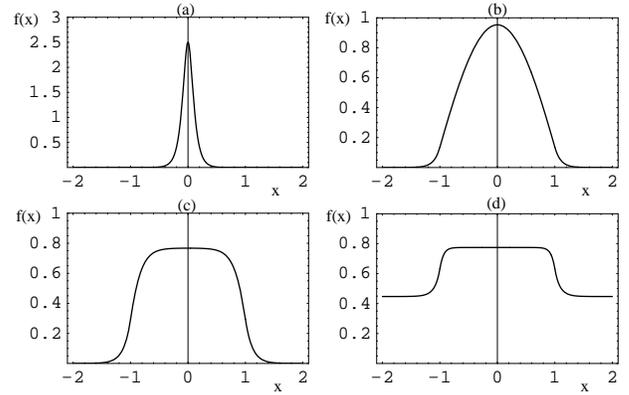,width=8.2cm}
\end{center}
\caption{
Tunable tunneling: shown are solutions for (a) negative scattering length, (b) no interactions between atoms, (c) positive scattering length and chemical potential less than the well depth, and (d) positive scattering length and chemical potential greater than the well depth.  Thus the tunneling and localization of the wavefunction is controlled via external experimental parameters.  A transition between the localized states shown in (a) - (c) and the non-normalizable, partially localized state shown in (d) can be characterized by the scattering length $a$.  For $10^4$ atoms of $^{85}$Rb in a trap of dimensions 100$\times$10$\times$10 $\mu$m$^3$ and depth 10 $\mu$K the transition occurs for $a\sim 14\,a_o$.}
\label{fig:5}
\end{figure}

The application of this work to tunable tunneling follows directly from the fact that $\eta\propto a N$.  In recent experimental developments it has been shown that the scattering length of certain Bose-condensed atoms, especially $^{85}$Rb, can be efficiently tuned over several orders of magnitude, including the sign~\cite{vogels1,cornish1}.  In Fig.~\ref{fig:5} the effect of this tuning is illustrated.  For $\eta < 0$, shown in panel (a), the tunneling can be made so small that the wavefunction is essentially unaffected by the potential.  In panel (b) the linear Schr\"odinger solution is shown, i.e. $\eta = 0$.  Panel (c) shows what happens when $\eta > 0$ and $\mu \leq V_o$.  The wavefunction is pushed far out into the barrier, as compared to the linear case in panel (b), but is still localized.  Finally, in panel (d) a partially localized state is shown.  $\eta$ is increased so much that $\mu > V_o$, and the wavefunction spills out over the top of the well and approaches a non-zero constant at infinity.  A well depth of $V_o=100$ has been used for the figure, but the general form of the transition is the same for any well.  Note that the tuning between panels (b) through (d) do not necessarily require a Feshbach resonance, as the density can be increased by evaporative cooling, which controls the temperature and therefore the number of atoms $N$ in the condensate, or by dynamically changing the trap potential.

In Fig.~\ref{fig:5} the localized states shown in panels (a) through (c) have a finite number of particles.  But as the partially localized state shown in panel (d) has a constant density as $x\rightarrow \pm \infty$ it requires an infinite number of particles.  This latter solution type, which has no analog in linear quantum mechanics, may be treated experimentally in two ways: either the entire system can be enclosed by a trap of infinite depth, as for example a shallow harmonic potential or an infinite square well potential with walls far from the finite square well potential, in which case it can be tuned by the scattering length or the density in the same manner as the localized solutions; or it can be controlled by the chemical potential $\mu > V_0$.  In present experiments the trap depth is effectively infinite and is usually idealized by a harmonic potential~\cite{ketterle1}.  Thus partially localized states have not been observed.  However, with the advent of BECs trapped in hollow blue-detuned laser beams capped on either end by laser light sheets of adjustable intensity~\cite{bongs1} it should now be possible to observe both localized and partially localized states.  Although $^{87}$Rb has been the atomic species used in hollow laser trap experiments, $^{85}$Rb could equally well be used~\cite{cornellcommunication1}.  Thus tunable tunneling can be studied with present BEC technology.

To explicate the usefulness of this application the physical scalings of Eqs.~(\ref{eqn:nls}) and~(\ref{eqn:potential}) are here worked out explicitly.  The quasi-one-dimensional Gross-Pitaevskii equation with tight transverse confinement and with all physical units is
\begin{eqnarray}
\left [-\frac{\hbar^2}{2m}\partial_{x_o}^{2}+\frac{4\pi \hbar^2 a N}{m L_y L_z}\mid\!f_o(x_o)\!\mid^{2} +\tilde{V}^{trap}(x)\,\right ]\,f_o(x_o) \nonumber \\
= \mu_o\,f_o(x_o)\, .
\label{eqn:nlso}
\end{eqnarray}
This equation is obtained by projection of the three-dimensional wavefunction onto the ground state in the transverse dimensions, which assumes an approximate separation of variables.  Details of this approximation are treated elsewhere~\cite{carr15,carr22,mueller1}.  In Eq.~(\ref{eqn:nlso}), box-like transverse confinement at $\pm L_y/2$ and $\pm L_z/2$ has been assumed.  Note that the experiment of Ref.~\cite{bongs1} is capable of producing these conditions~\cite{sengstockcommunication1,carr26}.  $a$ is the s-wave interaction length which characterizes binary atomic interactions and $N$ is the total number of atoms.  The proper scalings between Eq.~(\ref{eqn:nlso}) and Eqs.~(\ref{eqn:nls}) and~(\ref{eqn:potential}) are then $\mu_o = [\hbar^2/(2m\,l^2)]\, \mu$, $\tilde{V}_o= [\hbar^2/(2m\,l^2)]\, V_o$, $\eta = 8\pi a N l/(L_y L_z)$, $x_o = l\, x$, and $f_o = f/\sqrt{l}$.

For the case of $10^4$ atoms and a trap of dimensions $2l=100\mu$m and $L_y$, $L_z = 10 \mu$m, a trap of $\tilde{V}_o\sim 10 \mu$K depth would correspond to $V_o\sim 10^8$.  The transition between localized and partially localized states would occur, in this trap, at $\eta_0= 2.0\times 10^8$.  Assuming the use of a Feshbach resonance rather than a change in density to achieve the transition, the scattering length for $^{85}$Rb, normally $\sim -400\,a_o$~\cite{cornish1} , need only be tuned to $a\sim 14\, a_o$, well within the range of present experimental capabilities.

\section{Summary}
\label{sec:summary}

It has been demonstrated that the tunneling and localization of the macroscopic wavefunction which describes the mean field of Bose-Einstein condensates in a trap of finite depth is an external experimentally controllable parameter.  Specifically, the full set of localized and partially localized stationary states of the nonlinear Schr\"odinger equation with a finite square well potential have been presented.  Localized states are bound, and their tails approach zero at a large distance from the well.  Partially localized states are defined as states for which the tails approach a constant non-zero value at a large distance from the well but the density is still higher in the region of the well.  The latter have no analog in the linear Schr\"odinger equation.  Tuning between these different kinds of states could be studied with present BEC technology.

An obvious extension of these tunable tunneling results would be to consider the problem of a condensate in two adjacent finite square wells.  Analytic solutions of the NLS will also be available in this case~\cite{mahmud1}.  Tuning the nonlinearity $\eta$ can then increase or decrease the coupling between the spatially separated condensates.  The NLS will not provide a complete description as this coupling approaches zero~\cite{smerzi1,milburn1}, but the availability of analytic mean field solutions will provide the zeroth order nonlinear wave functions needed in models which include the correlations necessary to describe the decoupling and recoupling of localized condensates.


\acknowledgments

We thank Graeme Henkelman for inspirational discussions.  This work was supported by NSF Chemistry and Physics.


\appendix
\section{Proof that there are no nodeless trivial phase solutions}
\label{app:dn}

Consider localized, nodeless solutions for attractive nonlinearity, i.e. $\eta<0$, of the form
\begin{subeqnarray}
\label{eqn:app1}
  f_1(x)&=& A\,\text{sech}[k(x-b)]\, ,\\
  f_2(x)&=& A_2\,\text{dn}(k_2 x\mid m)\, ,\\
  f_3(x)&=& A\,\text{sech}[k(x+b)]\, ,
\end{subeqnarray}
where $A$, $k$, $k_2$, $b$, and $m$ are free parameters.  $f_1(x)$ and $f_3(x)$ have been chosen in a manner to preserve symmetry under the reflection operation $x\rightarrow -x$.  By the methods explicated in Sec.~\ref{subsubsec:att} the two systems of simultaneous equations which constrain the parameters are
\begin{subeqnarray}
\label{eqn:app2}
  A^2&=& -2k^2/\eta\, ,\\
  \mu&=& V_o - k^2\, ,\\
  A_2^2&=& -2mk_2^2/\eta\, ,\\
  \mu&=& - (2-m)k_2^2\, ,
\end{subeqnarray}
and
\begin{subeqnarray}
\label{eqn:app3}
&-(2-m)\beta^2 + \alpha^2 = \gamma^2\, ,\\
&\nonumber\\
&\alpha\,\text{sech}(\omega)=\beta\,\text{dn}(\beta\mid m) \, ,\\
&\nonumber\\
&\alpha^2\,\text{tanh}(\omega)\,\text{sech}(\omega)\nonumber\\
&=m\,\beta^2\,\text{cn}(\beta\mid m)\,\text{sn}(\beta\mid m) \, ,\\
&\nonumber\\
&\beta\,E(\beta\mid m)+\alpha\,e^{-\omega}\,\text{sech}(\omega)= -l\eta/4 \, ,
\end{subeqnarray}
where $\alpha\equiv k\,l$, $\beta\equiv k_2 l$, $\omega\equiv k(l+b)$, and $\gamma = \sqrt{V_o}l$.

Equations~(\ref{eqn:app3}) can be reduced to a set of two simultaneous equations in the parameters $\beta$ and $m$, again by the methods of Sec.~\ref{subsubsec:att}.  One of these equations, which must be satisfied, is
\begin{eqnarray}
\label{eqn:app4}
\sqrt{\lambda^2-\beta^2\,\text{dn}^2(\beta\mid m)}\,\text{dn}(\beta\mid m) \nonumber \\
-m\,\beta\,\text{cn}(\beta\mid m)\,\text{sn}(\beta\mid m) = 0\, ,
\end{eqnarray}
where $\lambda\equiv (2-m)\beta^2+\gamma^2$.  It shall be proven that this equation is false.

First note that $\beta$ and $\gamma$ are positive definite and $0<m<1$.  For Eq.~(\ref{eqn:app4}) to be true, $\text{cn}(\beta\mid m)\,\text{sn}(\beta\mid m)$ must be positive, so
\begin{eqnarray}
\label{eqn:app5}
0<\beta<K(m) \quad\text{or}\quad 2K(m)<\beta<3K(m).  
\end{eqnarray}
Then an absolute value sign may be imposed,
\begin{eqnarray}
\label{eqn:app6}
\sqrt{\lambda^2-\beta^2\,\text{dn}^2(\beta\mid m)}\,\text{dn}(\beta\mid m) \nonumber \\
-m\,\beta\,|\text{cn}(\beta\mid m)|\,|\text{sn}(\beta\mid m)| = 0\, .
\end{eqnarray}
Equation~(\ref{eqn:app5}) and the fact that $0<m<1$ imply that $\text{dn}(\beta\mid m)>|\text{cn}(\beta\mid m)|$.  Then
\begin{eqnarray}
\label{eqn:app7}
\sqrt{\lambda^2-\beta^2\,\text{dn}^2(\beta\mid m)} < m\,\beta\,|\text{sn}(\beta\mid m)|\, .
\end{eqnarray}
Square both sides and apply the elliptic identity $m\,\text{sn}^2(\beta\mid m)+\,\text{dn}^2(\beta\mid m)=1$.  This implies that $\lambda^2<\beta^2$.  Substitute back in $\lambda\equiv (2-m)\beta^2+\gamma^2$ and the condition becomes
\begin{eqnarray}
\label{eqn:app8}
(2-m)\beta^2+\gamma^2<\beta^2 \, .
\end{eqnarray}
This implies that $(2-m)<1$.  But $0<m<1$.  Therefore there are no nodeless trivial phase solutions in the finite square well for any set of well parameters.  Conversely, note that nodeless, dn-type solutions do exist in the case of a constant potential with periodic boundary conditions~\cite{carr16}.


\begin{references}
\bibitem[*]{byline} to whom correspondence should be addressed
\end{references}

%
%

%

%

%

%

\begin{table}
\caption{Limits of Jacobian elliptic functions and integrals.  The first three, i.e. sn, cn, and dn, are periodic solutions in the well while the $m=1$ limits of dn, ds, and ns, i.e. sech, csch, and coth, constitute the decaying tails of the wavefunction outside the well.  $4 K(m)$ is the periodicity and the elliptic integrals $K(m)$ and $E(m)$ both play a role in the system of equations which describe the solutions.}
\label{table:jacobian}
\begin{tabular}{ccc}
  & $m=0$ & $m=1$ \\
\tableline
$\text{sn}(u \mid m)$ & $\sin (u)$  & $\text{tanh}(u)$ \\
$\text{cn}(u \mid m)$ & $\cos (u)$  & $\text{sech}(u)$ \\
$\text{dn}(u \mid m)$ & 1         & $\text{sech}(u)$ \\
$\text{ds}(u \mid m)$ & $\text{csc}(u)$  & $\text{csch}(u)$ \\
$\text{ns}(u \mid m)$ & $\text{csc}(u)$  & $\text{coth}(u)$ \\
$K(m)$                  & $\pi/2$   & $\infty$       \\
$E(m)$                  & $\pi/2$   & 1              \\
\end{tabular}
\end{table}

\end{document}